\documentclass[lettersize,journal]{IEEEtran}
\usepackage{amsmath,amsfonts,stmaryrd,color}
\usepackage{array}
\usepackage[caption=false,font=normalsize,labelfont=sf,textfont=sf]{subfig}
\usepackage{textcomp}
\usepackage{stfloats}
\usepackage{url}
\usepackage{verbatim}
\usepackage{graphicx}
\usepackage{tikz}
\usepackage{pgfplots}
\usepackage{graphicx}
\usepackage{subcaption}
\pgfplotsset{compat=newest}
\usetikzlibrary{shapes,arrows,fit,positioning,calc}
\usetikzlibrary{plotmarks}
\usetikzlibrary{decorations.pathreplacing}
\usetikzlibrary{patterns}
\usetikzlibrary{chains,arrows,shapes,spy}
\usetikzlibrary{automata}
\usepackage{algpseudocode}

\begin{document}

\title{Iterative Detection and Decoding for Clustered Cell-Free Massive MIMO Networks }

\pdfinclusioncopyfonts=1
\author{\IEEEauthorblockN{Tonny Ssettumba\IEEEauthorrefmark{1}, Saeed Mashdour\IEEEauthorrefmark{1}, Lukas T. N. Landau\IEEEauthorrefmark{1}, Paulo B. da Silva \IEEEauthorrefmark{2}, and Rodrigo C. de Lamare\IEEEauthorrefmark{1}} \vspace{-0.5em}	
\vspace{-2.0em}}


\maketitle
\begin{abstract}
In this letter, we propose an iterative soft interference cancellation scheme for intra-cluster (ICL) and out-of-cluster (OCL) interference mitigation in user-centric clustered cell-free massive multiple-antenna networks. {We propose a minimum mean-square error receive filter with a novel modified parallel interference cancellation scheme to mitigate ICL and OCL interference. Unlike prior work, we model the OCL interference and devise a least squares estimator to perform OCL interference estimation.} 
An iterative detection and decoding scheme that adopts low-density parity check codes and incorporates the OCL interference estimate is developed. Simulations assess the proposed scheme against existing techniques in terms of bit error rate performance.
\end{abstract}

\begin{IEEEkeywords}
Out-of-cluster interference, intra-cluster interference, iterative interference cancellation. \vspace{-0.5em}
\end{IEEEkeywords}

\section{Introduction}
Cell-free massive multiple input multiple output (CF-mMIMO) systems differ from cellular architectures by deploying a large number of distributed access points (APs) in the coverage area. These APs operate to serve all users, regardless of their location, forming a distributed antenna system \cite{r0,r1,r1.5,rmmsecf,cesg,cfrs,cands}. The absence of cell boundaries in CF-mMIMO enables more efficient resource utilization and better interference management \cite{r2,jed,r3,r4} than in cellular networks \cite{jpba}. {However, the original version of CF-mMIMO systems is not scalable and is difficult to implement. Previous works proposed clustering techniques to make CF-mMIMO systems scalable and practical \cite{r5, r6}. }  

{However, as the number of UEs increases, clustered networks face interference from both intra-cluster (ICL) and out-of-cluster (OCL), making receiver design challenging \cite{r4,r5}. To effectively mitigate ICL and OCL interference, it is crucial to cancel signals of undesired UEs within the desired cluster and neighboring clusters, requiring interference estimation and the design of high-performance receivers.} Several receivers have been proposed for cellular, multi-input multi-output (MIMO) and massive MIMO systems \cite{r7}. However, cellular networks consider frequency reuse, which is absent in CF-mMIMO networks \cite{r1,r2,jed,r3}. {Several detection schemes for CF-mMIMO systems were studied in \cite{r2,jed,r3,r4,r8,r9}, even though the development of cost-effective receiver schemes remains an open challenge.} 
Therefore, there is need for the design of efficient cluster-based receivers that can cancel ICL and OCL interference. In the literature, there is no work so far on iterative ICL and OCL mitigation schemes for CF-mMIMO networks.

{Unlike the work in \cite{r91} that considers a signal-to-leakage-plus-noise (SLNR)-maximizing method for cancelling known interference in MIMO broadcast channels, we propose an iterative soft ICL and OCL interference cancellation scheme for the uplink of user-centric clustered CF-mMIMO networks \cite{oclidd}.} In particular, we propose a minimum mean-square error (MMSE) receive filter with a novel modified parallel interference cancellation (PIC) scheme for the proposed receiver in the presence of ICL and OCL interference and noise. Moreover, we model the OCL interference and devise a least-squares estimator (LSE) to perform OCL interference estimation. An iterative detection and decoding (IDD) scheme that adopts low-density parity-check (LPDC) codes and incorporates the OCL interference estimate is then presented. The proposed interference estimation and IDD scheme can be used with the proposed modified PIC and other interference cancellation strategies. Numerical results assess the proposed approaches against competing techniques. 

The remainder of this paper is organized as follows. The proposed system model, channel and interference estimation are presented in Section \ref{sys}. Section \ref{rec} details the proposed modified PIC receiver and the proposed IDD scheme. 
Simulation results and the associated discussion are presented in Section \ref{sim}. Finally, concluding remarks are given in \ref{conc}.

\textit{\textbf{Notation}}: We use lower and upper bold case symbols to represent vectors/matrices, respectively. The Hermitian transpose operator is denoted by $(\cdot)^{H}$.

\section{Proposed System Model, Channel and Interference Estimation} \label{sys}

\begin{figure}[htbp]
\centering
\includegraphics[width=7.5cm]{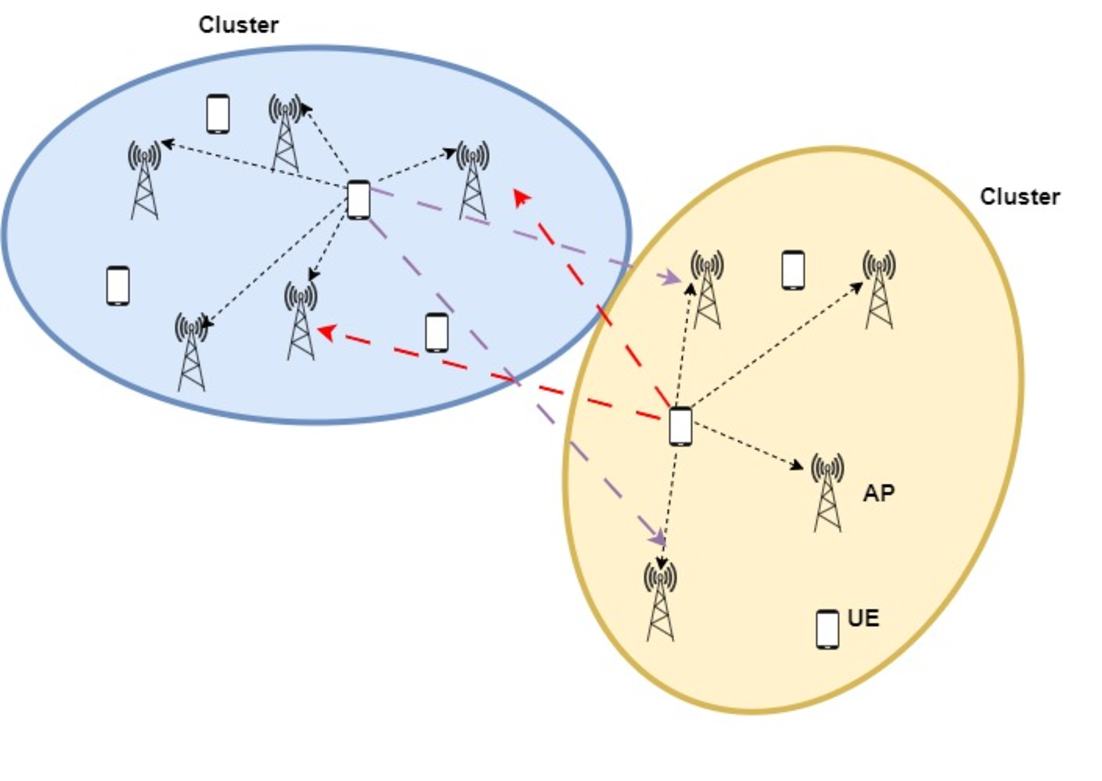}
\vspace{-1.75em}
\caption*{(a)}
\vspace{-0.1em}
\includegraphics[width=7.5cm]{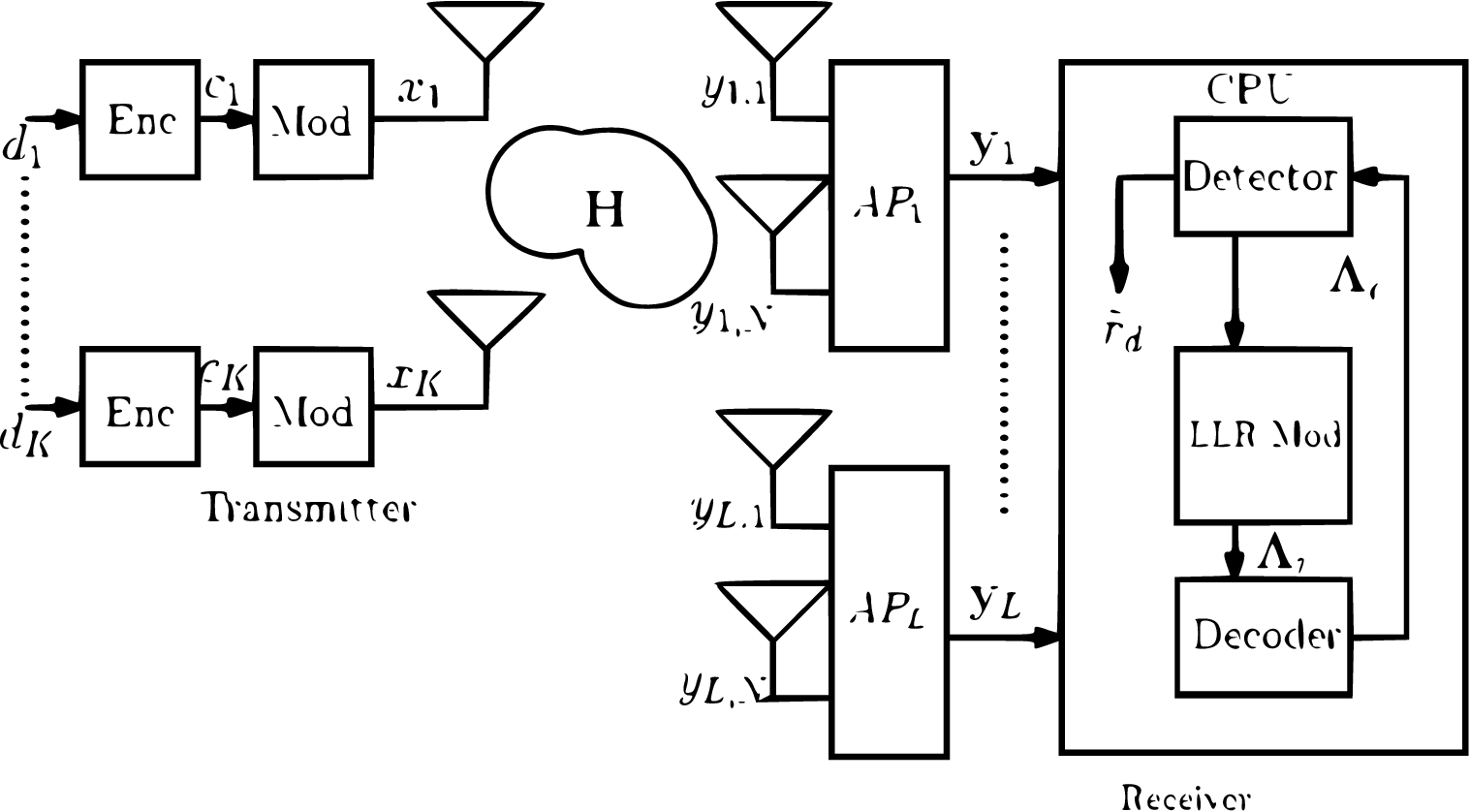}
\vspace{-1.75em}
\caption*{(b)}
\caption{(a) Schematic of the clustered cell-free systems. (b) Proposed IDD scheme with users and APs in a cluster.}
\label{fig_combined}
\end{figure}

The uplink of a user-centric clustered CF-mMIMO system \cite{r1.5} is considered  as shown in Fig. \ref{fig_combined}. Fig. \ref{fig_combined} (a) shows the desired  (dotted) links within a given cluster as well as OCL (dashed) links to each cluster. The proposed IDD scheme is presented in Fig. \ref{fig_combined} (b), where the transmitter comprises of $K$ single antenna user equipment (UEs), encoders which employ LDPC codes and quadrature phase-shift keying (QPSK) modulation or 16 quadrature-amplitude modulation (16-QAM). The system suffers from $M$ OCL sources of interference that are assumed to be identical to the UEs in terms of operation. The receiver consists of $L$ APs each equipped with $N$ antennas, and a central processing unit (CPU) which comprises a  detector, a log-likelihood ratio (LLR) computing module (LLR Mod) and an LDPC decoder. The detector computes a symbol estimate $\tilde{r}_{d}$ and sends it to the LLR Mod which computes the intrinsic LLRs $\Lambda_{i}$. The LLRs obtained are then sent to the decoder which computes extrinsic LLRs $\Lambda_{e}$. The exchange of LLRs between the detector and the decoder is done in an iterative fashion to improve performance.

\subsection{Channel Estimation for the Serving Users}

To estimate the channels, $K$ mutually orthogonal $\tau_{p}\geq K$ length pilot signals are assumed to be assigned to UE $k$ denoted by $\phi_{k}$, with $\left\| \phi_{k} \right\|^{2}=1$. The received signal during the pilot phase at AP $l$ is given by \cite{r7}
\begin{align}  \mathbf{Y}_{l}=\sqrt{p\tau_{p}}\mathbf{H}_{l}\mathbf{\Phi}^{H}+\mathbf{G}_{l}\mathbf{S}^{H}+\mathbf{N}_{l}
\end{align}
where $p$, $\mathbf{H}_{l} \in \mathbb{C}^{N\times K}$, $\mathbf{\Phi}\in \mathbb{C}^{\tau_{p}\times K}$, $\mathbf{G}_{l}\in \mathbb{C}^{N\times M}$,$\mathbf{S}\in \mathbb{C}^{\tau_{p}\times M}$, and $\mathbf{N}_{l}\in \mathbb{C}^{N\times \tau_{p}}$ is the transmit power of each UE, the channel matrix between UEs and $l$-th AP,
the matrix of pilot symbols,
the channel matrix between the OCL interference and the $l$-th AP, the transmitted signal matrix of OCL interference, and the receiver noise at the $l$-th AP, respectively. The LSE is then applied to compute the channel estimate:
\begin{align}
\hat{\mathbf{H}}_{l}=\frac{1}{\sqrt{p\tau_{p}}}\mathbf{Y}_{l}\mathbf{\Phi},
\end{align}
where $\hat{\mathbf{H}}_{l}$ is tht channel estimate, $\tau_{p}$ is the pilot length. 
\subsection{Interference Estimation}

Unlike prior work on estimation of out-of-system channels \cite{r7}, we consider the estimation of OCL interference in user-centric clustered CF-mMIMO networks. To estimate the OCL interference, the received signal at each AP is first pre-processed to reduce the impact of the channel components of the serving UEs from them, i.e $\left\{\mathbf{H}_{l}, l=1,...,L\right\}$. This preprocessed signal is the residual signal \cite{r7} given by
\begin{align}\label{residual_signal}
\mathbf{Z}_{l}=\mathbf{Y}_{l}-\sqrt{p\tau_{p}}\hat{\mathbf{H}}_{l}\mathbf{\Phi}^{H}.
\end{align}
After some mathematical manipulations and following similar steps as in \cite{r7}, \eqref{residual_signal} can be written as
\begin{align}   \mathbf{Z}_{l}=\left(\mathbf{G}_{l}\mathbf{S}^{H}+\mathbf{N}_{l}\right)\mathbf{P}^{\perp},
\end{align}
where $\mathbf{P}^{\perp}=\left(\mathbf{I}-\mathbf{\Phi}\mathbf{\Phi}^{H}\right)$ is the projection onto the orthogonal complement of the pilot matrix $\mathbf{\Phi}$.

{\bf Remarks:} Note that the projection matrix is not invertible, and thus the OCL interfering signals cannot be completely estimated. Therefore, only its components spanned by $\mathbf{P}^{\perp}$ can be estimated \cite{r7}. Thus, to make meaningful the estimates of $\mathbf{S}$ from $\mathbf{Z}_{l}$, the condition $\tau_{p}>K+M$ should be verified. Otherwise, the residual matrices in all APs will be zero \cite{r7}. The projection matrix $\mathbf{P}^{\perp}$ can be decomposed as
\begin{align}
    \mathbf{P}^{\perp}=\mathbf{	\Psi}\mathbf{	\Psi}^{H},
\end{align}
where $\mathbf{\Psi}\in \mathbb{C}^{\tau_{p}\times (\tau_{p}-K)}$ is a tall matrix that satisfies the orthogonality property $\mathbf{	\Psi}^{H}\mathbf{	\Psi}=\mathbf{I}$, which yields the economy-size singular value decomposition (SVD) of $\mathbf{P}$. For clusters of variable size, the proposed techniques will require dynamic pilot lengths. To estimate and mitigate ICL and OCL interference, the excess number of antennas in the cluster APs will need to be sufficient to allow the components spanned by the projection matrix ${\mathbf P}^{\perp}$ to be estimated and the number of pilots must be greater than $K$ and $M$, i.e., $\tau_p>K+M$. The projected signal component is given by
\begin{align}  \bar{\mathbf{S}}=\mathbf{\Psi}^{H}\mathbf{S}.
\end{align}

In order to estimate $\hat{\bar{\mathbf{S}}}$  of $\bar{\mathbf{S}}$, the residual signal is first despread by projecting it on $\mathbf{\Psi}$ as follows 
\begin{align}
\mathbf{Z}_{l}\mathbf{\Psi}&=\left(\mathbf{G}_{l}\mathbf{S}^{H}+\mathbf{N}_{l}\right)\mathbf{\Psi},\notag\\&
    =\mathbf{G}_{l}\bar{\mathbf{S}}^{H}+\mathbf{N}^{'}_{l},
\end{align}
where $\mathbf{N}^{'}_{l}=\mathbf{N}_{l}\mathbf{\Psi}$. With the signals processed by each AP and the channel estimates, the OCL interference was obtained by sequential accumulation of Gramians $\mathbf{\Psi}^H \mathbf{Z}^H\mathbf{Z}\mathbf{\Psi}=(\mathbf{Z}_{l}\mathbf{\Psi})^H(\mathbf{Z}_{l}\mathbf{\Psi})$ and the centralized scheme of \cite{r7}. In this scheme, the processed signals from all APs are collected at the CPU and the following least squares problem is solved:
\begin{align}   \min_{\mathbf{G},~\bar{\bar{\mathbf{S}}}}||\mathbf{Z}\mathbf{\Psi}-\mathbf{G}\bar{\mathbf{S}}^{H}||_{F},
\end{align}
where $\mathbf{Z}=\left[ \
\mathbf{Z}^{T}_{1},....,\mathbf{Z}^{T}_{L} \right]$, $\mathbf{G}=\left[ \
\mathbf{G}^{T}_{1},....,\mathbf{G}^{T}_{L} \right]$, $\left\| \cdot \right\|_{F}$ denotes the Frobenius norm of the argument. The solution to the problem is obtained by taking the best rank-1 approximation of $\mathbf{Z}\mathbf{\Psi}$ using the SVD. The estimated values of the matrices $\mathbf{G}$ and $\mathbf{S}$ are the left and right singular matrices of $\mathbf{Z}\mathbf{\Psi}$ scaled by the largest singular value \cite{r7}, which requires $O(KNL)$.

\vspace{-2mm}
\subsection{Uplink Data Transmission}

The received signal at the $l$th AP after estimating the channels of the UEs and the OCL interference is given by
\begin{align}   \mathbf{y}_{l}=\mathbf{H}_{l}\mathbf{x}+\mathbf{G}_{l}\mathbf{s}+\mathbf{n}_{l}, \label{rec_sig}
\end{align}
where $\mathbf{H}_{l}\in \mathbb{C}^{N\times K}$, $\mathbf{x}\in \mathbb{C}^{K\times 1}$, $\mathbf{G}_{l}\in \mathbb{C}^{N\times M}$, $\mathbf{s}\in \mathbb{C}^{M\times 1}$, $\mathbf{n}_{l}\in \mathbb{C}^{N\times 1}$, is the channel matrix of the UEs in the system, the transmitted signals of the UEs in the system, the channel matrix of the OCL interference, the transmitted symbol for the OCL interference and additive white Gaussian noise (AWGN) with zero mean and unit variance, respectively.

\section{Proposed Iterative Detection and Decoding} \label{rec}
In this section, we present the proposed IDD scheme and the design of the proposed linear and modified PIC receivers.

\subsection{Proposed Iterative Receiver Design}

The proposed soft receiver aims to cancel the ICL and OCL interference. The demodulator comprises an MMSE filter followed by a modified PIC scheme, which is modified to cancel both $K$ ICL and $M$ OCL interfering signals. The receiver first computes the $j$th UE symbol mean $\bar{s}_{j}$ to obtain soft estimates  \cite{r2,r3,r10}. The symbol mean is defined as $\mathbb{E}\left \{ s_{j} \right \}= \bar{s_{j}}$ and is given by
\begin{align}\label{expectation_sym}
\bar{s}_{j}=\sum_{s\in\mathcal{A}}s P(s_{j}=s),
\end{align}
where $\mathcal{A}$ is the set of complex constellations.
The a priori probability of the extrinsic LLRs is given by \cite{r2,r3,r10}
\begin{align}\label{aprior_prob}
	 P(s_{j}=s)=\prod_{l=1}^{M_{c}}\lbrack 1+\exp(-s^{b_{l}}\Lambda_{i}(b_{(j-1)M_{c}+l}))\rbrack^{-1},
\end{align}
where $\Lambda_{i}(b_{i})$ is the extrinsic LLR of the $i$-th bit calculated by the LDPC decoder from the previous iteration, and $s^{b_{l}}\in (+1,-1)$ denotes the $l$-th bit of symbol $s$.
The variance of the $j$-th UE symbol is calculated as \cite{r2,r3,r10}
\begin{align}\label{variex}
	\sigma^{2}_{j} =\sum_{s\in \mathcal{A}}|s-\bar{s}_{j}|^{2}P(s_{j}=s). 
	\end{align} 
After channel estimation and interference estimation, the received signal in \eqref{rec_sig} can be re-written as
\begin{equation}
  \mathbf{y}_{l}=\left[ \begin{matrix}
\hat{\mathbf{H}}_{l}& \hat{\mathbf{G}}_{l} 
\end{matrix} \right]\left[ \begin{matrix}
\mathbf{x} \\\bar{\mathbf{s}}
\end{matrix} \right]+\mathbf{n}_{l}
\end{equation}
Let us define $\mathbf{A}_{l}=\left[ \begin{matrix}
\hat{\mathbf{H}}_{l}& \hat{\mathbf{G}}_{l} 
\end{matrix} \right]\in \mathbb{C}^{N\times (K+M)}$ and $\mathbf{r}=\left[ \begin{matrix}
\mathbf{x} \\\bar{\mathbf{s}}
\end{matrix} \right]\in \mathbb{C}^{ (K+M)\times1 }$, which allows the received signal at the $l$-th AP be described by
\begin{align}   \mathbf{y}_{l}=\mathbf{A}_{l}\mathbf{r}+\mathbf{n}_{l}
\end{align}
For centralized processing, the received signal at the CPU is
\begin{align} \label{sig_cpu} \mathbf{y}=\mathbf{A}\mathbf{r}+\mathbf{n},
\end{align}
where $\mathbf{A}=\left[ \
\mathbf{A}^{T}_{1},....,\mathbf{A}^{T}_{L} \right]$ is the matrix consisting of all the channels from the serving UEs and the OCL interfering signals.
The received signal in \eqref{sig_cpu} can be decomposed as
\begin{align}   
\mathbf{y}=\mathbf{a}_{\text{d}}r_{\text{d}}+\sum_{i=1,i\neq \text{d}}^{K+M}\mathbf{a}_{\text{i}}r_{\text{i}}+\mathbf{n}, \label{sig_dec}
\end{align}
where $\mathbf{a}_{d}$ and $r_{d}$, are the channel vector and transmitted symbol of the desired signal, the second term denotes the interference signal from the other $K$-1 UEs and $M$ OCL interferers and the third term denotes the noise. The detected signal for the desired UE after the modified PIC is given by
\begin{align}  \tilde{r}_{d}=\mathbf{w}^{H}_{d}\biggl(\mathbf{a}_{\text{d}}r_{\text{d}}+\sum_{i=1,i\neq \text{d}}^{K+M}\mathbf{a}_{\text{i}}\left( r_{\text{i}}-\bar{r}_{\text{i}} \right)+\mathbf{n}\biggr),\label{det_sig}  
\end{align}
where $\mathbf{w}_{d}$ is the receive combining vector. The proposed modified PIC alters the upper limit of the above summation to allow cancellation of $M$ OCL interfering signals, whereas the proposed linear receiver does not employ interference cancellation but considers both ICL and OCL in the covariance matrix used to compute the MMSE receive filter. The design of a receive filter $\mathbf{w}_{d}$ aims at minimizing the error between the desired user signal $r_{d}$ and the estimated signal $\tilde{r}_{d}$. The optimization problem to minimize the  mean square error (MSE)  between $\tilde{r}_{d}$ and $r_{d}$ is formulated as
\begin{align}
 \text{MSE}=\mathbb{E}\left\{ \left| r_{\text{d}}-\tilde{r}_{\text{d}} \right|^{2} \right\}. \label{mse1}
\end{align}
By substituting  \eqref{det_sig} into \eqref{mse1}, yields the MSE given by
\begin{align} \label{MSE}
&\text{MSE}=-\mathbf{w}^{H}_{\text{d}}\rho_{\text{d}}\mathbf{a}_{\text{d}}\\&+\notag\mathbf{w}^{H}_{\text{d}} \biggl[ \rho_{\text{d}}\mathbf{a}_{\text{d}}\mathbf{a}^{H}_{\text{d}}+\sum_{i=1,i\neq\text{d}}^{K+1}\mathbf{a}_{\text{i}}\mathbb{E}\left\{ \left| r_{\text{i}}-\tilde{r}_{\text{i}} \right|^{2} \right\}\mathbf{a}^{H}_{\text{i}}+\sigma^{2}_{n}\mathbf{I}_{NL} \biggr] \mathbf{w}_{\text{d}}.
\end{align}
By differentiating \eqref{MSE} with respect to (w.r.t) $\mathbf{w}^{H}_{\text{d}}$ and equating to zero, the MMSE receive filter is given by
\begin{align}
\mathbf{w}_{\text{d}}=&\rho_{\text{d}}\biggl[ \rho_{\text{d}}\mathbf{a}_{\text{d}}\mathbf{a}^{H}_{\text{d}}+ \vspace{-0.5em} \sum_{i=1,i\neq\text{d}}^{K+M}\mathbf{a}_{\text{i}}\mathbb{E}\left\{ \left| r_{\text{i}}-\tilde{r}_{\text{i}} \right|^{2} \right\}\mathbf{a}^{H}_{\text{i}}+\sigma^{2}_{n}\mathbf{I}_{NL} \biggr]^{-1}  \mathbf{a}_{\text{d}}.   
\end{align}
The main factors affecting the performance of the receiver are the number of OCL interferers, the parameter $\mathbb{E}\{ \left| r_{\text{i}}-\tilde{r}_{\text{i}} \right|^{2}\}$ which consists of entries computed in \eqref{variex}, the number of UEs $K$, as well as the number of APs $L$. The impact of these parameters on the performance are discussed in detail in the simulations section. Modified successive interference cancellation (SIC) and list-based  receivers would alter the modified PIC by changing the number of cancellation elements in the summation of \eqref{det_sig}. A Gaussian approximation of the receiver output is employed and detailed next.

\subsection{IDD Scheme with Gaussian Approximation }\label{IDD}

The received signal at the output of the filter in \eqref{det_sig}, contains the desired symbol, ICL and OCL interference as well as  the noise. We use similar assumptions given  in \cite{r2,r3,r10,jidf,spa,mfsic,mbdf,dfcc,did,bfidd,1bitidd,listmtc,detmtc,msgamp1,msgamp2,dynovs,comp} to approximate the $\tilde{r}_{d}$ as a Gaussian output given by 
\begin{align}\label{Gauaprroxi}  \tilde{r}_{d}=\mu_{\text{d}}r_{d}+z_{\text{d}},
\end{align}
By comparing \eqref{Gauaprroxi} with  
$\mu_{\text{d}}=\mathbf{w}^{H}_{d}\mathbf{a}_{d}$, the parameter $z_{\text{d}}$ is a zero-mean AWGN variable  given by
\begin{equation}   z_{\text{d}}=\mathbf{w}^{H}_{d}\biggl(\sum_{i=1,i\neq \text{d}}^{K+M}\mathbf{a}_{\text{i}}\left( r_{\text{i}}-\bar{r}_{\text{i}} \right)+\mathbf{n}\biggr)
\end{equation}
The variance $\sigma^{2}_{z}=\mathbb{E}\{z_{\text{d}}z^{*}_{\text{d}}\}$ of $z_{\text{d}}$ is given by 
\begin{align}  \sigma^{2}_{z}&=\mathbf{w}^{H}_{\text{d}}\biggl[ \sum_{i=1,i\neq\text{d}}^{K+M}\mathbf{a}_{\text{i}}\mathbb{E}\left\{ \left| r_{\text{i}}-\tilde{r}_{\text{i}} \right|^{2} \right\}\mathbf{a}^{H}_{\text{i}}+\sigma^{2}_{n}\mathbf{I}_{NL} \biggr]\mathbf{w}_{\text{d}}
\end{align}
The extrinsic LLR computed by the soft MMSE receiver for the $l$-th bit $l\in\left\{1,2,...,M_{c}\right\}$ of  symbol $r_{d}$  \cite{r2,r3} is given by
\begin{equation}
\begin{split}
\label{LLR_COMP}
   &\Lambda_{e}\left ( b_{(d-1)M_{c}+l} \right )=\notag\\&\frac{\log P\left ( b_{(d-1)M_{c}+l}=+1 |\tilde{r}_{d}\right)}{\log P\left ( b_{(d-1)M_{c}+l}=-1| \tilde{r}_{d}\right)}-\frac{\log P\left ( b_{(d-1)M_{c+1}}=+1 \right )}{\log P\left ( b_{(d-1)M_{c+1}}=-1 \right )} \\&
    =\log\frac{\sum _{s\in \mathcal{A}^{+1}_{l}}F\left ( \tilde{r}_{d}|s \right )P\left (s \right )}{\sum _{s\in \mathcal{A}^{-1}_{l}}F\left ( \tilde{r}_{d}|s \right )P\left (s \right )}-\Lambda_{i}\left ( b_{(d-1)M_{c}+l} \right )\notag,
 \end{split}
 \end{equation}
 where Bayes's rule is applied to obtain the last equality of \eqref{LLR_COMP}. The $2^{Mc-1}$  hypothesis set for which the $l$-th  bit is $+1$ is denoted by the parameter  $\mathcal{A}^{+1}_{l}$.  The a priori probability $P(s)$ is obtained from \eqref{aprior_prob}. The likelihood function $F(\tilde{r}_{d}|s)$ is approximated as \cite{r2,r3} 
    \begin{equation}\label{llfn}
        F\left ( \tilde{r}_{d}|s \right )\simeq\frac{1}{\pi\sigma^{2}_{z}}\exp\left (-\frac{1}{\sigma^{2}_{z}} |\tilde{r}_{d}-\mu_{d}s|^{2} \right ).
    \end{equation}

 The decoder and the proposed detectors exchange soft beliefs in an iterative manner. The performance of the traditional sum-product algorithm (SPA) is negatively impacted by the tangent function, particularly in the error-rate region. Thus, we adopt the box-plus SPA in this work as used in our earlier papers \cite{r2,r3} since it produces less complex approximations as compared to the former. The decoder comprises two steps namely: Single parity check (SPC) stage and the repetition stage. The LLRs sent from check node $(CN)_{J}$ to variable node $(VN)_{i}$  are computed as 
\begin{equation}
        \Lambda_{j \longrightarrow i}= {\oplus}_{i^{'}\in N(j) / i} \Lambda_{i^{'\longrightarrow j}},
\end{equation}
where $\oplus$ denotes the pairwise ``box-plus" operator given by 
\begin{align}
 \Lambda_{1}\oplus \Lambda_{2}=&\log\left ( \frac{1+e^{\Lambda_{1}+\Lambda_{2}}}{e^{\Lambda_{1}}+e^{\Lambda_{2}}} \right ),\\\notag
   =&\mathrm{sign}(\Lambda_{1})\mathrm{sign}(\Lambda_{2})\min(\left | \Lambda_{1} \right |,\left | \Lambda_{2} \right |)\\&\notag+\log\left ( 1+e^{-\left |\Lambda_{1}+\Lambda_{2}  \right |} \right )-\log\left (1+e^{-\left |\Lambda_{1}-\Lambda_{2}  \right |}  \right ).
\end{align}
The LLR from $VN_{i}$ to $CN_{j}$ is given by
\begin{align}
 \Lambda_{i\longrightarrow j} = \Lambda_{i}+\sum_{j^{'} \in N(i)\backslash j}\Lambda_{j^{'}\longrightarrow i}, 
\end{align}
 where the parameter $\Lambda_{i}$ denotes the LLR at $VN_{i}$, ${j^{'}\in N(i)\backslash j}$ means that all CNs connected to $VN_{i}$ except $CN_{j}$. 
 The cost of the proposed IDD scheme for OCL interference cancellation is $O(N^2LK)$, which is comparable to standard schemes. 
 
 \section{Simulation Results} \label{sim}
 
 The performance of the proposed approaches is assessed in this section. The SNR used in the simulations is defined by
\begin{equation}
    SNR=\frac{\sum_{l=1}^{L}(\mathbf{H}_{l}~\mathrm{diag}\left(\bf {\rho}\right)\mathbf{H}^{H}_{l})}{\sigma_{w}^{2} NL K}.
\end{equation}
The simulation parameters are in Table \ref{simpara}.
\begin{table}[h!]
\renewcommand*{\arraystretch}{1.5}
\begin{footnotesize}
\caption{Simulation Parameters.}
\vspace{-1.25em}
\begin{center}
\begin{tabular}{|p{5cm}|p{2cm}|}
\hline
\textbf{Parameter} & \textbf{{Value}} \\
\hline
Codeword length $(n)$ & $512$  \\
\hline
Parity Check bits ($n-k$) & $256$\\
\hline
Message bits $(k)$ & $128$\\
\hline
Code rate $R$  & $\frac{1}{2}$\\
\hline
$\tau_{u}, \tau_{p}, \tau_{c}$ & $190, 10, 200$ \\
\hline
Maximum decoder iterations & $10$ \\
 \hline
 Signal power $\rho$ & $1~\mathrm{W}$  \\
 \hline
Maximum decoder iterations & $10$ \\
   \hline
Number of channel realizations  & $10000$  \\
\hline
$L$, $N$, $K$ and $M$ & $32$, $4$, $8$ and $4$    \\
\hline
\end{tabular}
\label{simpara}
\end{center}
\end{footnotesize}
\end{table}

\textbf{Network setup}: We consider a cell-free cluster  with squared dimensions of $D= 1$ km. We also consider an LDPC code and message passing decoding \cite{memd,vfap} for the proposed IDD scheme. The  channels between OCL interferers and APs are generated randomly by assuming them to follow block fading.
The QPSK and QAM-16 modulation schemes are used and the LS fading coefficients are obtained according to the 3rd Generation Partnership Project (3GPP) Urban Microcell model in \cite{r1} given by
\begin{align}     \beta_{k,l}\left[\mathbf{\mathrm{dB}}\right]=-30.5-36.7\log_{10}\biggl(\frac{d_{kl}}{1 m}\biggr)+\Upsilon_{kl},
  \end{align}
 where $d_{kl}$ is the distance between the $k$-th UE and $l$-th AP, $\Upsilon_{kl}\sim\mathcal{N}\left(0, 4^{2}\right)$ is the shadow fading \cite{r1}.

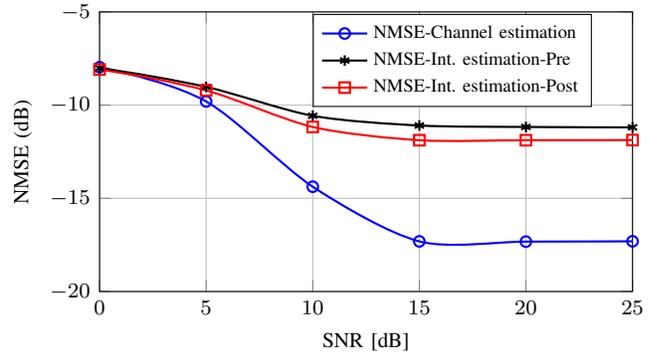
\begin{figure}[!htbp]
\centering
%
%
%
\usetikzlibrary{positioning,calc}

\definecolor{mycolor1}{rgb}{0.00000,1.00000,1.00000}%
\definecolor{mycolor2}{rgb}{1.00000,0.00000,1.00000}%

\definecolor{mustard}{rgb}{0.92941,0.69020,0.12941}%

\definecolor{newpurple}{rgb}{0.5, 0 ,1}%

\definecolor{darkblue}{rgb}{0, 0.4470, 0.7410}

\pgfplotsset{every axis label/.append style={font=\footnotesize},
every tick label/.append style={font=\footnotesize},
every plot/.append style={ultra thick} 
}

\begin{tikzpicture}[font=\footnotesize] 

\begin{axis}[%
name=mse,
width  = 0.8\columnwidth,
height = 0.42\columnwidth,
scale only axis,
xmin  = 0,
xmax  = 25,
xlabel= {SNR [dB]},
xmajorgrids,
ymin=-20,
ymax=-5,
ylabel={NMSE (dB)},
ymajorgrids,
legend entries={NMSE-Channel estimation,
                NMSE-Int. estimation-Pre,
			NMSE-Int. estimation-Post,
				},
legend style={fill=white, fill opacity=0.8, draw opacity=1,
text opacity =1,at={(0.4,0.66)}, anchor= south west,draw=black,fill=white,legend cell align=left,font=\scriptsize}
]

\addlegendimage{smooth,color=blue,solid, thick, mark=o,
y filter/.code={\pgfmathparse{\pgfmathresult-0}\pgfmathresult}}
\addlegendimage{smooth,color=black,solid, thick, mark=asterisk,
y filter/.code={\pgfmathparse{\pgfmathresult-0}\pgfmathresult}}
\addlegendimage{smooth,color=red,solid, thick, mark=square,
y filter/.code={\pgfmathparse{\pgfmathresult-0}\pgfmathresult}}

\addplot+[smooth,color=blue,solid,thick, every mark/.append style={solid} ,mark=o,
y filter/.code={\pgfmathparse{\pgfmathresult-0}\pgfmathresult}]
  table[row sep=crcr]{%
-5	-7.309467721099234\\
0	-7.962893332523017\\
5	-9.802881798477785\\
10 -14.377395516916058\\
15 -17.310802467456040\\
20	-17.323535167227070\\
25	-17.306024596384205\\
};

\addplot+[smooth,color=black,solid, thick, every mark/.append style={solid} ,mark=asterisk,
y filter/.code={\pgfmathparse{\pgfmathresult-0}\pgfmathresult}]
  table[row sep=crcr]{%
-5	-7.555700366312389	\\
0	-8.01771822288337\\
5	-9.02201530790056\\
10  -10.571680835699952\\
15  -11.0940735540031\\
20	-11.1720094151540\\
25	-11.1941184122840\\
};

\addplot+[smooth,color=red,solid, thick, every mark/.append style={solid} ,mark=square,
y filter/.code={\pgfmathparse{\pgfmathresult-0}\pgfmathresult}]
  table[row sep=crcr]{%
-5	-7.645700366312389	\\
0	-8.091771822288337\\
5	-9.211208530790056\\
10  -11.171680835699952\\
15  -11.881840735540031\\
20	-11.882006094151540\\
25	-11.874101184122840\\
};


\end{axis}
\end{tikzpicture}
\vspace{-0.5em}
\caption{NMSE versus SNR  $L=32$, $N=4$, $K=8$, $M=4$  for OCL interference estimation and channel estimation.}
 \label{figJ11} 
\end{figure}

Fig. \ref{figJ11} presents the normalized mean square error (NMSE) for the interference estimation and channel estimation. We consider the interference estimation before (pre) and after (post) channel estimation. It can be observed that for both cases the NMSE reduces as we approach higher SNR values up to the error floor region. For channel estimation this can be further reduced by using longer pilots to provide more accurate estimation. The interference estimation after channel estimation was adopted as it performed slightly better than before channel estimation. Further improvements in performance can be obtained by using more advanced estimators. Accurate estimation of the interference is key in the IDD cancellation step as it improves the accuracy and efficiency of the receiver. 

 \begin{figure}[!htbp]
\centering
%
%
%
\usetikzlibrary{positioning,calc}

\definecolor{mycolor1}{rgb}{0.00000,1.00000,1.00000}%
\definecolor{mycolor2}{rgb}{1.00000,0.00000,1.00000}%

\definecolor{mustard}{rgb}{0.92941,0.69020,0.12941}%

\definecolor{newpurple}{rgb}{0.5, 0 ,1}%

\definecolor{darkblue}{rgb}{0, 0.4470, 0.7410}

\pgfplotsset{every axis label/.append style={font=\footnotesize},
every tick label/.append style={font=\footnotesize},
every plot/.append style={ultra thick} 
}

\begin{tikzpicture}[font=\footnotesize] 

\begin{axis}[%
name=mse,
width  = 0.55\columnwidth,
height = 0.395\columnwidth,
scale only axis,
xmin  = -5,
xmax  = 25,
xlabel= {SNR [dB]},
xmajorgrids,
ymin=0.001,
ymax=0.1,
ymode=log,
ylabel={BER},
ymajorgrids,
legend entries={IDD=1,
					IDD=2,
     			IDD=3,
				},
legend style={fill=white, fill opacity=0.8, draw opacity=1,
text opacity =1,at={(0.05,0.05)}, anchor= south west,draw=black,fill=white,legend cell align=left,font=\scriptsize}
]

\addlegendimage{smooth,color=black,dotted, thick, mark=x,
y filter/.code={\pgfmathparse{\pgfmathresult-0}\pgfmathresult}}
\addlegendimage{smooth,color=red,dashed, thick, mark=square,
y filter/.code={\pgfmathparse{\pgfmathresult-0}\pgfmathresult}}
\addlegendimage{smooth,color=blue,solid, thick, mark=o,
y filter/.code={\pgfmathparse{\pgfmathresult-0}\pgfmathresult}}
\addplot+[smooth,color=black,dotted,thick, every mark/.append style={solid} ,mark=x,
y filter/.code={\pgfmathparse{\pgfmathresult-0}\pgfmathresult}]
  table[row sep=crcr]{%
-5	0.116650439453125	\\
0	0.079459912109375\\
5	0.056075927734375 \\
10 0.042263037109375\\
15 0.033648193359375 \\
20	0.02987080078125\\
25	0.02744609375\\
};

\addplot+[smooth,color=red,dashed, thick, every mark/.append style={solid} ,mark=square,
y filter/.code={\pgfmathparse{\pgfmathresult-0}\pgfmathresult}]
  table[row sep=crcr]{%
-5	0.105730810546875	\\
0	0.063828369140625\\
5	0.036205419921875\\
10  0.020134130859375\\
15  0.01093212890625 \\
20	0.005659912109375\\
25	0.003327392578125\\
};


\addplot+[smooth,color=blue,solid, thick, every mark/.append style={solid} ,mark=o,
y filter/.code={\pgfmathparse{\pgfmathresult-0}\pgfmathresult}]
  table[row sep=crcr]{%
-5	0.105646240234375	\\
0	0.063635546875\\
5	0.035876708984375 \\
10 0.019773095703125\\
15  0.0105904296875\\
20	0.005294580078125\\
25	0.002965576171875\\
};
\end{axis}
\end{tikzpicture}
\vspace{-1.05em}
\caption{\small BER versus SNR  $L=32$, $N=4$, $K=8$, $M=4$ for the system with ICL and OCL interference, QPSK, and a varying number of modified PIC IDD iterations.}
 \label{figJ12} 
\end{figure}
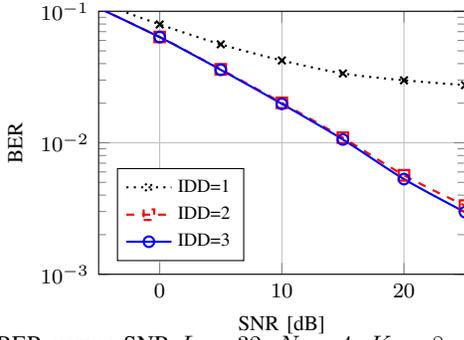

The BER performance versus SNR for a varying number of IDD iterations for the system that suffers from both ICL and OCL interference is shown in Fig. \ref{figJ12}. Note that as the number of iterations increases from the first to the second iteration, there is a significant reduction in BER. From the second to the third iteration, there is a marginal reduction in BER in the high SNR regime. This improvement is due to the fact that as the number of IDD iterations is increased, extra a posterior information is exchanged between the decoder and the detector which improves the quality of the soft beliefs used by the interference cancellation schemes. 

\begin{figure*}[!htbp]
\centering
\input{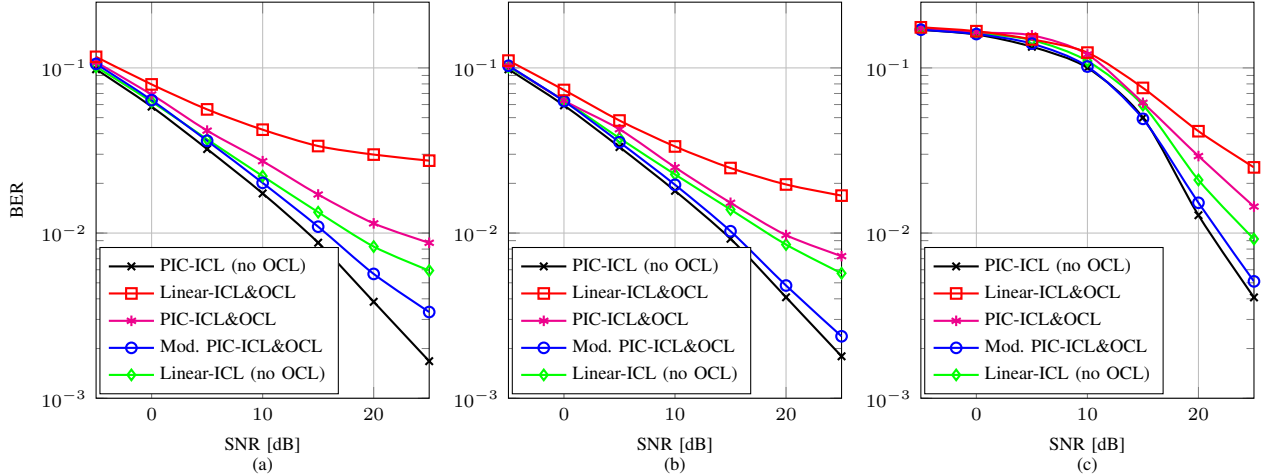}
\vspace{-0.925em}
\caption{BER versus SNR  for (a) $L=32$, $N=4$, $K=8$, $M=4$, QPSK, (b)  $L=32$, $N=4$, $K=8$, $M=2$, QPSK and (c) $L=32$, $N=4$, $K=8$, $M=2$, 16-QAM for the clustered network.}
 \label{figJ13} 
\end{figure*}
Fig. \ref{figJ13} shows the BER versus SNR for the system with OCL and ICL interference. Note that the system that suffers only from ICL interference achieves lower BER values than those with OCL and ICL interference. Moreover, the proposed modified PIC receiver achieves lower BER values than the linear receiver with both ICL and OCL, the PIC receiver with both ICL and OCL and the linear receiver with ICL and no OCL as it can cancel more effectively multiple streams of interference. Figs. \ref{figJ13} (a) and \ref{figJ13} (b) evaluate the impact of reducing the number of OCL interference sources from 4 to 2. The curves in Fig. \ref{figJ13} (b) achieve a lower BER than those in Fig. \ref{figJ13} (a) for all receivers with OCL interference. Fig. \ref{figJ13} (c) assesses the impact of 16-QAM modulation on the proposed IDD scheme, which shows the same performance hierarchy as those in Fig. \ref{figJ13} (a) and (b). The iterative processing becomes more accurate with ICL and OCL interference cancellation of the modified
PIC receiver, which lowers the level of interference experienced by the IDD scheme.

\vspace{-2.5mm}
 \section{Concluding Remarks} \label{conc}
 
 In this paper, we have investigated soft interference cancellation schemes for user-centric clustered CF-mMIMO networks in the presence of ICL and OCL interference. We developed MMSE receive filters, proposed a modified PIC receiver and devised an LSE to perform OCL interference estimation. An IDD scheme that adopts LDPC codes and incorporates the OCL interference estimate and the modified PIC receiver was presented. The results showed that the proposed techniques outperform competing techniques and approach the performance of the system without OCL interference.

\end{document}